\documentclass[fleqn,usenatbib]{mnras}

\usepackage{newtxtext,newtxmath}

\usepackage[T1]{fontenc}

\usepackage{graphicx}	
\usepackage{amsmath}	

\title[X--ray, UV, \& optical variability in NGC\,6814]{Characterizing X--ray, UV, and optical variability in NGC\,6814 using high--cadence \textit{Swift} observations from a 2022 monitoring campaign}

\author[A. G. Gonzalez et al.]{
A. G. Gonzalez$^{1}$\thanks{E-mail: adam.gonzalez@smu.ca (AGG)},
L. C. Gallo$^{1}$,
J. M. Miller$^{2}$,
E. S. Kammoun$^{3,4}$,
A. Ghosh$^{1}$,
and B. A. Pottie$^{1}$
\\
$^{1}$Department of Astronomy and Physics, Saint Mary's University, 923 Robie Street, Halifax, NS, B3H 3C3, Canada\\
$^{2}$Department of Astronomy, University of Michigan, 1085 South University Avenue, Ann Arbor, MI, 48109-1107, USA\\
$^{3}$IRAP, Universit\'e de Toulouse, CNRS, UPS, CNES, 9 Avenue du Colonel Roche, BP 44346, 31028 Toulouse Cedex 4, France\\
$^{4}$INAF – Osservatorio Astrofisico di Arcetri, Largo Enrico Fermi 5, I-50125 Firenze, Italy}

\date{Accepted XXX. Received YYY; in original form ZZZ}

\pubyear{2023}

\begin{document}
\label{firstpage}
\pagerange{\pageref{firstpage}--\pageref{lastpage}}
\maketitle

\begin{abstract}
We present the first results of a high--cadence \textit{Swift} monitoring campaign ($3-4$ visits per day for $75$ days) of the Seyfert 1.5 galaxy NGC\,6814 characterizing its variability throughout the X--ray and UV/optical wavebands. Structure function analysis reveals an X--ray power law ($\alpha=0.5^{+0.2}_{-0.1}$) that is significantly flatter than the one measured in the UV/optical bands ($\langle\alpha\rangle\approx1.5$), suggesting different physical mechanisms driving the observed variability in each emission region. The structure function break--time is consistent across the UV/optical bands ($\langle\tau\rangle\approx2.3~\mathrm{d}$), suggesting a very compact emission region in the disc. Correlated short time--scale variability measured through cross--correlation analysis finds a lag--wavelength spectrum that is inconsistent with a standard disc reprocessing scenario ($\tau\propto\lambda^{4/3}$) due to significant flattening in the optical wavebands. Flux--flux analysis finds an extremely blue AGN spectral component ($F_{\nu}\propto\lambda^{-0.85}$) that does not follow a standard accretion disc profile ($F_{\nu}\propto\lambda^{-1/3}$). While extreme outer disc truncation ($R_{\mathrm{out}}=202\pm5~r_g$) at a standard accretion rate ($\dot{m}_{\mathrm{Edd}}=0.0255\pm0.0006$) may explain the shape of the AGN spectral component, the lag--wavelength spectrum requires more modest truncation ($R_{\mathrm{out}}=1,382^{+398}_{-404}~r_g$) at an extreme accretion rate ($\dot{m}_{\mathrm{Edd}}=1.3^{+2.1}_{-0.9}$). No combination of parameters can simultaneously explain both results in a self--consistent way. Our results offer the first evidence of a non--standard accretion disc in NGC\,6814.
\end{abstract}




\section{Introduction} \label{sec:introduction}


Active galactic nuclei (AGNs) host a central supermassive black hole (SMBH) that is actively accreting material from its surrounding environment, producing the most continuously luminous source of emission in the Universe. The extreme compactness of the SMBH and its environment, however, renders this region unresolved by current telescopes. Indirect methodologies, such as the analysis of multi--wavelength flux variability, have therefore been developed to study the physical processes at work in this extreme region.

X--ray variability studies have revealed that close to the SMBH exists an extremely hot ($T\sim10^9~\mathrm{K}$), optically thin ($\tau<1$), compact \citep[$r\sim10~r_g$; e.g.][]{Alston+2020,GalloGonzalezMiller2021} X--ray corona. This corona illuminates an optically thick, geometrically thin disc of accreting material that surrounds the SMBH \citep{PringleRees1972,ShakuraSunyaev1973,NovikovThorne1973,LyndenBellPringle1974}, which itself produces thermal UV emission at the inner region and optical emission further out due to a radial temperature profile that follows $T \propto r^{-3/4}$ \citep{ShakuraSunyaev1973}. The coronal and disc emission comprise the high--energy spectral energy distribution (SED) of AGNs, and act as a photoionising continuum for material at larger distances.

Optical variability studies have measured the size of the so--called broad line region (BLR) in AGNs, responsible for the production of broad optical emission lines, through the reverberation mapping technique \citep{BlandfordMcKee1982,Peterson1993}. By measuring time delays between correlated optical continuum and emission line variations, and assuming any measured time delay between the two is solely due to light travel time effects, the size of the BLR can be estimated as $10^{3}-10^{5}~r_g$ \citep[e.g.][]{GalloMillerCostantini2023}. Combining this result with the width of the corresponding emission line, a SMBH mass estimate can also be made, which has yielded an ever growing sample of reverberation--based SMBH mass estimates \citep[e.g.][]{BentzKatz2015}.

On smaller scales, time delays between the X--ray and UV/optical emission in AGNs have been used to estimate the size of the accretion disc. X--rays illuminating the accretion disc will be both reprocessed and emitted by the disc \citep[i.e. so--called X--ray reflection;][]{GeorgeFabian1991} as well as absorbed by the disc, thereby heating it and producing thermal UV/optical emission. By correlating the X--ray variability with UV/optical continuum emission in various energy bands, corresponding time delays can be measured that can therefore probe both the location of the X--ray corona \cite[e.g.][]{Kammoun+2021_sample} and the size of the accretion disc \cite[e.g.][]{McHardy+2023}. 

NGC\,6814 \citep[$z=0.00522$;][]{Springob+2005} is a Seyfert 1.5 \citep{VeronCettyVeron2006} hosted by a nearly face--on grand--design spiral galaxy \citep[see Figure 1 of][]{Bentz+2019}\footnote{Image also available at \href{https://www.nasa.gov/image-feature/goddard/2016/hubble-spies-a-spiral-snowflake}{https://www.nasa.gov/image-feature/goddard/2016/hubble-spies-a-spiral-snowflake}}. Its SMBH mass is estimated to be $M_{\mathrm{BH}}=\left(1.85\pm0.35\right)\times10^{7}~M_{\odot}$ based on H$\beta$ measurements \citep{Bentz+2009}, though a range of $M_{\mathrm{BH}}\approx\left(0.55-6.1\right)\times10^{7}~M_{\odot}$ have also been reported using a variety of H and He emission lines \citep{Bentz+2010}, with dynamical BLR modelling finding $M_{\mathrm{BH}}=2.6^{+1.9}_{-0.9}\times10^{6}~M_{\odot}$ \citep{Pancoast+2014}. Recently, we presented evidence of a partial X--ray eclipse in NGC\,6814 that lasted $\sim42~\mathrm{ks}$ \citep{GalloGonzalezMiller2021}. Combined with the duration of ingress and egress (both $\sim13.5~\mathrm{ks}$), we were able to estimate the distance to the obscuring material as $r \approx 2700~r_g$ and its density as $n_e \approx 10^{10}~\mathrm{cm}^{-3}$, consistent with a BLR origin \citep[e.g.][]{Rees+1989}. The X--ray corona was also estimated to have a diameter of $\sim25~r_g$, consistent with X--ray reverberation measurements \citep[e.g.][]{Alston+2020}. New work by \cite{Pottie+2023} has revealed that the obscuring medium is likely inhomogeneous, with denser clumps embedded in an ionised halo.

Here we present the results of a high--cadence ($3-4$ visits per day) multi--wavelength monitoring campaign of NGC\,6814 with the \textit{Swift} \citep{Gehrels+2004} observatory carried out from 28 August to 10 November 2022 (75 days). We outline our data processing steps in Section \ref{sec:data}, with all data products and results of the various analyses performed presented in Section \ref{sec:analysis}. We discuss our findings in Section \ref{sec:discussion} and conclude in Section \ref{sec:conclusions}.

\section{Observations \& Data Reduction} \label{sec:data}
Below we describe our processing steps for data collected with the X--ray Telescope \citep[XRT;][]{Burrows+2005} and the UltraViolet/Optical Telescope \cite[UVOT;][]{Roming+2005} onboard \textit{Swift}.

The XRT was operated in Photon Counting (PC) mode for the entire 2022 observing campaign. We processed the XRT data using the online XRT Product Builder tool \citep{Evans+2007,Evans+2009}, which is a pipe--line that produces data products that have been treated for pile--up, dead pixels, and vignetting. Using the per--snapshot binning scheme, we extracted the X--ray light curve in four energy bands of $0.3-10$ (broad), $0.3-1$ (soft), $1-4$ (medium), and $4-10$ (hard) $\mathrm{keV}$ to explore any possible energy--dependence of the forthcoming results. In the end, we found all of our results to be consistent across all four energy bands, and therefore present only the broad band X--ray results here to be concise (references to the \textit{X} waveband throughout thus refer to $0.3-10~\mathrm{keV}$). We also extracted the time--averaged XRT spectrum for our campaign using the same tool, including all events with valid grades (i.e. $0-12$).

The UVOT was operated in imaging mode for the entire observing campaign, where for each pointing we used the \textit{UVW2}, \textit{UVM2}, \textit{UVW1}, \textit{U}, \textit{B}, and \textit{V} filters. We extracted the source and background fluxes from each image in the standard way as outlined in the UVOT Data Analysis guide\footnote{\href{https://www.swift.ac.uk/analysis/uvot/mag.php}{https://www.swift.ac.uk/analysis/uvot/mag.php}}. We used the \textsc{uvotsource} task from the \textsc{ftools} package included in \textsc{heasoft} version 6.28 using a source region of radius $5''$ centred on the source position as available on the NASA/IPAC Extragalactic Database (NED) and a nearby off--source background region of $15''$. To account for so--called UVOT dropouts, in which the source flux is observed to be significantly decreased compared to its neighbours due to bad pixels on the UVOT detector, we followed the steps outlined in the Appendix of \cite{Edelson+2015}. This iterative filtering procedure removes 2, 2, 9, 0, 1, and 0 points from the final \textit{UVW2}, \textit{UVM2}, \textit{UVW1}, \textit{U}, \textit{B}, and \textit{V} light curves, respectively.

\begin{table}
	\caption{Light curve properties for each waveband of the 2022 observing campaign of NGC\,6814. The columns are: (1) waveband / filter, (2) number of observations, (3) median cadence, and (4) fractional variability.}
	\centering
	\begin{tabular}{cccc}
		\hline
		Waveband & $N_{\mathrm{obs}}$ & $\Delta t_{\mathrm{med}} \left[\mathrm{d}\right]$ & $F_{\mathrm{var}}$ \\
		\hline
		\textit{X}  & $260$ & $0.257$ & $0.42\pm0.02$ \\
		\textit{W2} & $251$ & $0.260$ & $0.25\pm0.01$ \\
		\textit{M2} & $247$ & $0.264$ & $0.22\pm0.01$ \\
		\textit{W1} & $247$ & $0.263$ & $0.176\pm0.008$ \\
		\textit{U}  & $255$ & $0.259$ & $0.155\pm0.007$ \\
		\textit{B}  & $252$ & $0.259$ & $0.078\pm0.004$ \\
		\textit{V}  & $250$ & $0.263$ & $0.037\pm0.003$ \\
		\hline
		\label{tab:obslog}
	\end{tabular}
\end{table}

A summary of the final light curve products is given in Table \ref{tab:obslog}.

\section{Analysis \& Results} \label{sec:analysis}

\subsection{Light curves} \label{subsec:lightcurves}

\begin{figure*}
	\begin{center}
		\scalebox{1.0}{\includegraphics[width=0.49\linewidth]{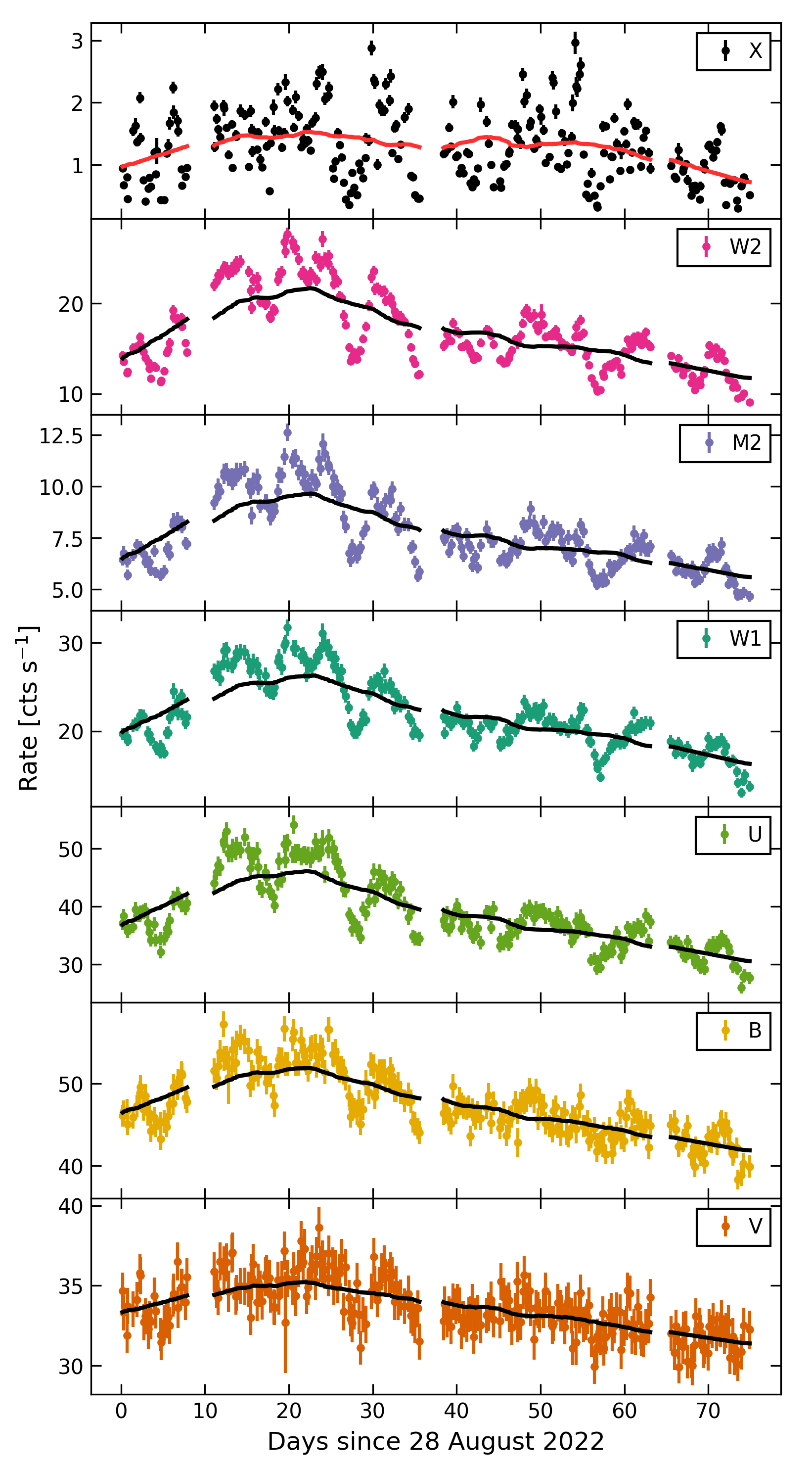}}
        \scalebox{1.0}{\includegraphics[width=0.49\linewidth]{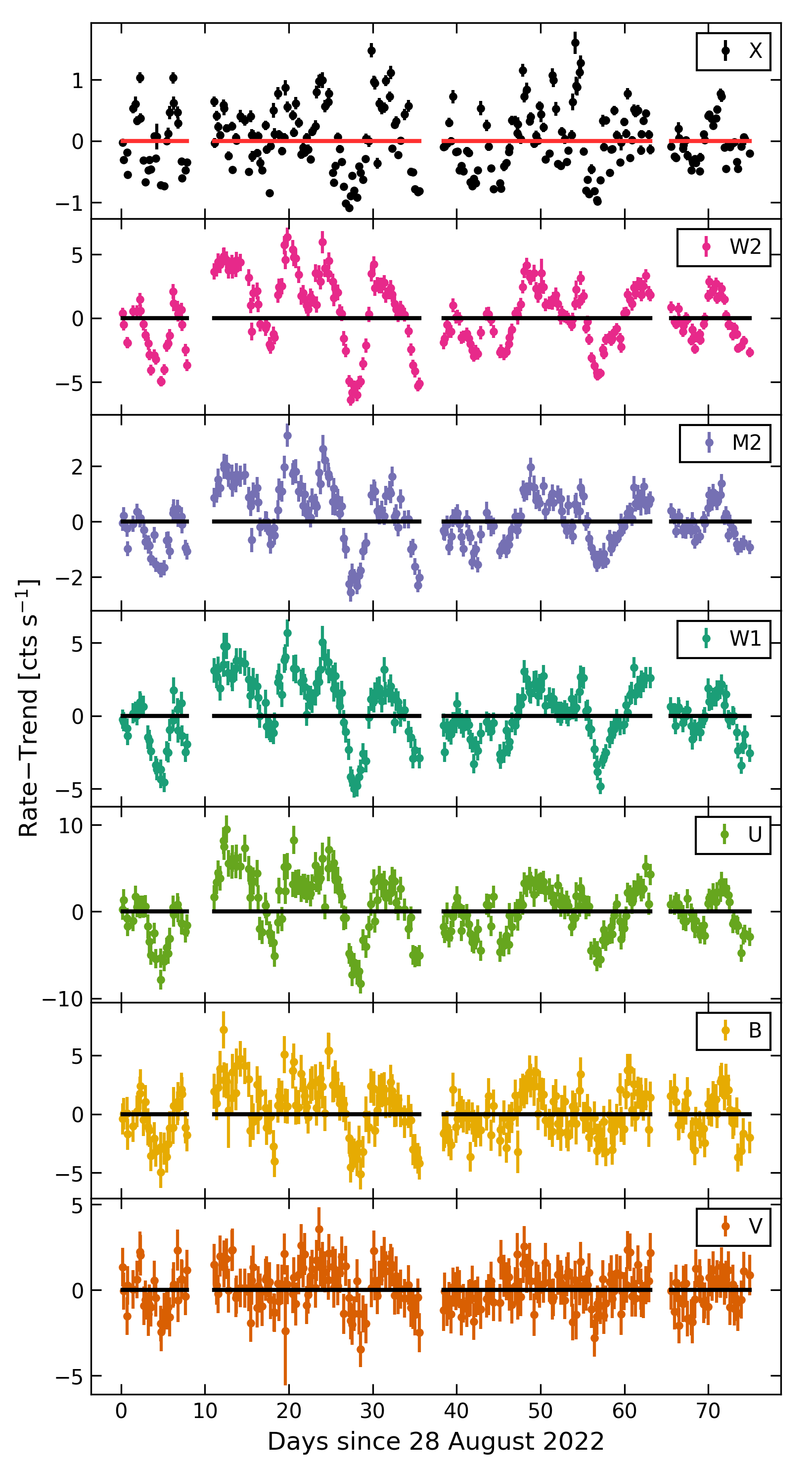}}
		\caption{\textit{Left}: Light curves for the 2022 observing campaign of NGC\,6814. UVOT dropouts have been removed (see text for details). Solid curves in each panel represent the Savitzky--Golay $21.5~\mathrm{d}$ smoothed trend (see Appendix \ref{app:detfiltwidth} for details). \textit{Right}: The de--trended (data minus smoothed trend in left panels) light curves. This procedure renders the \textit{V} de--trended light curve indistinguishable from a constant count rate ($\chi^{2}_{\nu}=1.03$), indicating that it is dominated by long--term variations not due to the reprocessing of X--rays by the accretion disc. All other de--trended light curves are inconsistent with a constant ($\chi^{2}_{\nu}>2$).}
		\label{fig:lightcurves}
	\end{center}
\end{figure*}

In Figure \ref{fig:lightcurves} (left panels) we present the light curves obtained during our 2022 observing campaign of NGC\,6814. All wavebands exhibit large amplitude, rapid variability that decreases at longer wavelengths. A common long--term trend is apparent in all of the UV/optical data, while the long--term X--ray variability appears distinctly different by comparison, suggesting unrelated or disconnected physical processes responsible for the long--term variability in each wavelength regime. 

To quantify the long--term variability, we compute the fractional variability ($F_{\mathrm{var}}$) following \cite{Edelson+2002} as:
\begin{equation} \label{eqn:fvar}
    F_{\mathrm{var}} = \sqrt{\frac{s^{2}-\langle\sigma_{\mathrm{err}}^{2}\rangle}{\langle X \rangle ^{2}}},
\end{equation}
where $X$ is the light curve count rate, $s^{2}$ is its variance, and $\sigma_{\mathrm{err}}^2$ is the square of its standard error. Its corresponding uncertainty is then computed as:
\begin{equation} \label{eqn:fvarerr}
    \sigma_{F_{\mathrm{var}}} = \frac{1}{F_{\mathrm{var}}} \sqrt{\frac{1}{2N}} \frac{s^{2}}{\langle X \rangle ^{2}},
\end{equation}
where $N$ is the number of light curve data points over which $F_{\mathrm{var}}$ is computed. Here, we compute the fractional variability over the entirety of each light curve. The results are given in Table \ref{tab:obslog}, where a clear trend of decreased variability with increased wavelength is observed. 

In order to isolate the short--term variations, which may be associated with the reprocessing of X--rays in the accretion disc, from long--term variations, which may be due to changes in the behaviour of the accretion flow itself, we de--trended the light curves using a Savitzky--Golay filter width of $21.5~\mathrm{d}$, which was determined by minimizing the \textit{W2} auto--correlation function variance (see Appendix \ref{app:detfiltwidth} for details). Moving forward, any reference to de--trended data has had the aforementioned trend line subtracted from the unaltered data, as shown in Figure \ref{fig:lightcurves} (right panels). The forthcoming analysis is conducted on both the unaltered and de--trended light curves, which will be presented alongside one another throughout. While our interpretation is based solely on the de--trended results they are largely independent of the usage of unaltered or de--trended data.

\subsection{Structure function} \label{subsec:structurefunction}
Due to the gappy, unevenly sampled light curves that comprise our data set, we are unable to implement Fourier--transform based techniques, such as the power density spectrum, without binning the data and therefore losing information on the shortest time--scales. To explore the distribution of power in our light curves as a function of time--scale ($\tau$) we instead compute the structure function (SF) according to the method of \cite{CollierPeterson2001} as:
\begin{equation} \label{eqn:sf}
    S\left(\tau\right) = \frac{1}{N\left(\tau\right)} \sum_{i<j} \left[X\left(t_i\right)-X\left(t_j\right)\right]^2,
\end{equation}
where $X\left(t\right)$ is the light curve count rate at time $t$, and the summation is performed over those data pairs $N\left(\tau\right)$ that satisfy $\tau=t_j-t_i$. We then subtract the noise variance ($2\sigma^{2}_{\mathrm{noise}}$) and normalize by the signal variance ($2\sigma^{2}_{\mathrm{signal}}$) such that the SF converges to $1$ with increasing time--scale. Within some range of time--scales where the variability is correlated the SF will take on a power law shape ($S\propto\tau^{\alpha}$) before flattening after some characteristic break--time. Similarly to the power density spectrum, the slope ($\alpha$) and break--time ($\tau_{\mathrm{break}}$) of the SF can inform us about the nature of the physical mechanism driving the observed variability \citep[e.g.][]{Gallo+2018}.

\begin{figure}
	\begin{center}
		\scalebox{1.0}{\includegraphics[width=\linewidth]{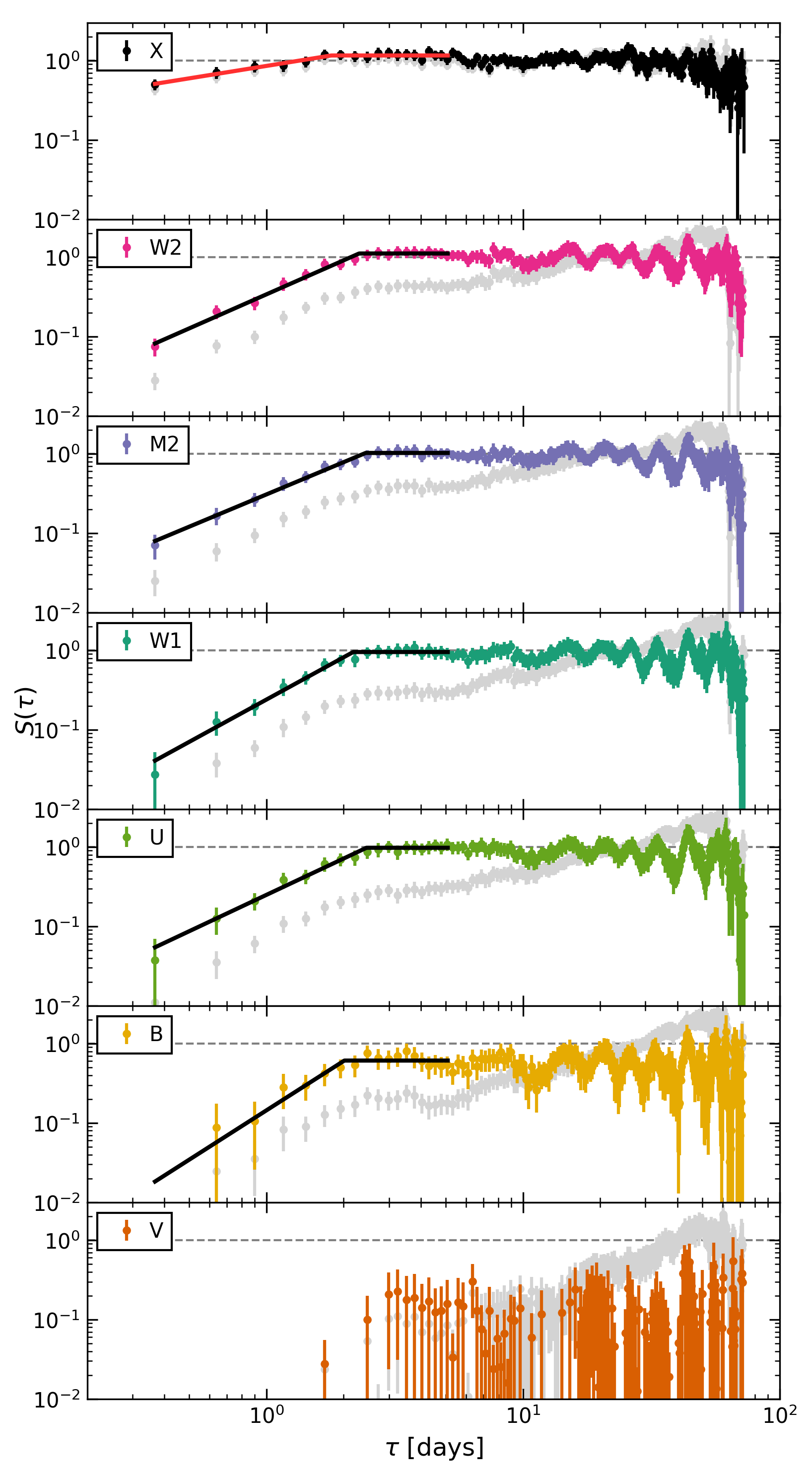}}
		\caption{Structure functions for each unaltered (grey) and de--trended (coloured) light curve. The horizontal dashed grey lines represent $S\left(\tau\right)=1$. Solid curves represent the median broken power law model plotted over the range of data points used to perform the fits. The \textit{V} band data, once de--trended, are invariant and thus are not fit, though we show them for comparison.}
		\label{fig:structurefunctions}
	\end{center}
\end{figure}

The SFs of the unaltered and de--trended light curves are shown in Figure \ref{fig:structurefunctions}. We have fit the de--trended SFs using a simple broken power law model\footnote{We note that we do not include an additional free--to--vary constant in our model as we have already subtracted the noise variance component.} such that $S\left(\tau<\tau_{\mathrm{break}}\right)=N\tau^{\alpha}$ and $S\left(\tau\geq\tau_{\mathrm{break}}\right)=N\tau_{\mathrm{break}}^{\alpha}$. To estimate parameters and their credible intervals we implement a Markov Chain Monte Carlo (MCMC) sampler via the \textsc{emcee} \citep{EMCEE} \textsc{python} package. In our MCMC runs we used a Gaussian likelihood function with uniform priors for each parameter, ran 30 walkers across the parameter space, burned the first $1,000$ iterations, and computed relevant parameter values ($50^{\mathrm{th}}$ percentile, i.e. median) and credible intervals ($16^{\mathrm{th}}$ and $84^{\mathrm{th}}$ percentiles) from the posterior distributions formed from a total of $10,000$ iterations per walker. We note that only the portion of the SF for which $\tau<5.1~\mathrm{d}$ was fit, as above this value the de--trended SFs begin to oscillate around $S\left(\tau\right)=1$. Furthermore, at $\tau\approx5.1~\mathrm{d}$ the unaltered SFs begin to deviate significantly from the nearly identical trend (albeit shifted in normalization) observed in the de--trended SFs. 

\begin{table}
	\caption{Structure function normalization ($N$), slope ($\alpha$), and break--time ($\tau_{\mathrm{break}}$) median values from the MCMC broken power law fits to the 2022 de--trended light curves, as well as reduced $\chi^{2}$ fit statistic ($\chi^{2}_{\nu}$). Parameter uncertainties represent the $68~\mathrm{per~cent}$ credible interval. A $\left(*\right)$ indicates a structure function that was not fit.}
	\centering
	\begin{tabular}{ccccc}
		\hline
		Waveband & $N$ & $\alpha$ & $\tau_{\mathrm{break}}$ [days] & $\chi^{2}_{\nu}$ \\
		\hline
		\textit{X}  & $0.86^{+0.07}_{-0.05}$ & $0.5^{+0.2}_{-0.1}$ & $1.8^{+0.5}_{-0.4}$ & $0.17$ \\
		\textit{W2} & $0.34\pm0.03$          & $1.4^{+0.2}_{-0.1}$ & $2.3\pm0.2$         & $0.16$ \\
		\textit{M2} & $0.31\pm0.03$          & $1.4^{+0.2}_{-0.1}$ & $2.4^{+0.3}_{-0.2}$ & $0.14$ \\
		\textit{W1} & $0.24\pm0.03$          & $1.8^{+0.3}_{-0.2}$ & $2.2^{+0.3}_{-0.2}$ & $0.16$ \\
		\textit{U}  & $0.25\pm0.03$          & $1.5^{+0.3}_{-0.2}$ & $2.5\pm0.3$         & $0.16$ \\
		\textit{B}  & $0.14^{+0.06}_{-0.07}$ & $2.1^{+1.5}_{-0.7}$ & $2.0^{+0.4}_{-0.3}$ & $0.40$ \\
		\textit{V}  & $\left(*\right)$       & $\left(*\right)$    & $\left(*\right)$    & $\left(*\right)$\\
		\hline
		\label{tab:sfparams}
	\end{tabular}
\end{table}

The median model curves are plotted alongside the data in Figure \ref{fig:structurefunctions}, with parameter values and credible intervals listed in Table \ref{tab:sfparams}. We find that the UV/optical SF slopes and break--times exhibit no significant wavelength dependence and are broadly consistent with their respective weighted means of $\langle\alpha\rangle \approx 1.46$ and $\langle\tau_{\mathrm{break}}\rangle \approx 2.30~\mathrm{d}$. These results suggest a common emission region in the disc is responsible for the observed variability across all UV/optical wavebands. The X--ray data, however, exhibit a significantly flatter SF slope than the mean UV/optical results (at the $6.3\sigma$ level) with a comparable SF break--time (within $\sim1\sigma$). While the similar X--ray and UV/optical break--times may suggest that the variability observed in both is produced in a similar emission region, the significant difference in SF slopes between the two wavelength regimes suggests a different physical process drives the observed variability in each.

\subsection{Interpolated cross--correlation function} \label{sec:iccf}
To explore the correlated variations between light curves in different wavebands, we computed the interpolated cross--correlation function\footnote{Our results do not change significantly when using other cross--correlation analysis techniques such as, for example, the discrete correlation function method of \cite{EdelsonKrolik1988}.} (ICCF; \citealt{GaskellSparke1986,GaskellPeterson1987,WhitePeterson1994}). This procedure computes the cross--correlation function (CCF) between two unevenly sampled light curves, $a\left(t\right)$ and $b\left(t\right)$, by linearly interpolating between the data points in each curve. The ICCF between two light curves shifted by a time--lag ($\tau$) is computed as:
\begin{equation} \label{eqn:iccf}
    \mathrm{ICCF}\left(\tau\right) = \frac{\langle\left[a\left(t\right)-\langle{a}\rangle\right]\left[b\left(t+\tau\right)-\langle{b}\rangle\right]\rangle}{\sigma_{a}~\sigma_{b}},
\end{equation}
where $\sigma$ denotes the standard deviation of the corresponding light curve count rate. In practice, our ICCFs were computed following the modifications by \cite{Edelson+2019}: (i) we did not compute the ICCF for pairs outside of the time series, (ii) all quantities in Equation \ref{eqn:iccf} were computed locally (i.e. over only the subset in each time series that overlaps for a given time--lag) rather than globally, and (iii) all ICCFs were computed using `two--way' interpolation by averaging over the individual results obtained by first interpolating $a\left(t\right)$ and leaving $b\left(t\right)$ gappy, then by interpolating $b\left(t\right)$ leaving $a\left(t\right)$ gappy. In this work we used the \textit{W2} light curve as the reference band, computing all ICCFs and thus all time--lags relative to it. We interpolated the light curves using a sampling rate a factor of $10$ smaller than the median cadence (see Table \ref{tab:obslog}).

To evaluate the significance of any peaks observed in the ICCFs, we simulated $N=1,000$ realizations of the \textit{W2} light curve using the light curve generation method of \cite{Emmanoulopoulos+2013} by estimating the slope of the power density spectrum ($\beta$) to be $\beta_{\textit{W2}}\approx\alpha_{\textit{W2}}+1=2.4$ and using the observed flux distribution as the input probability density function. We computed the ICCF for each realization, producing a distribution of noisy ICCFs. We then computed the $1^{\mathrm{st}}$ and $99^{\mathrm{th}}$ percentiles from the ensemble of noisy ICCFs at each time--lag, and only considered ICCF peaks from the data to be a significant result if they laid below/above these thresholds. 

Each ICCF was found to contain a single significant positive peak near $\tau\approx0~\mathrm{d}$. We therefore used the subset of each ICCF $>99^{\mathrm{th}}$ percentile from the corresponding simulations and computed the centroid ($\tau_{\mathrm{cent}}$) as:
\begin{equation} \label{eqn:centroid}
    \tau_{\mathrm{cent}} = \frac{\sum{\tau\times\mathrm{ICCF}\left(\tau\right)}}{\sum{\mathrm{ICCF}\left(\tau\right)}}.
\end{equation}
To estimate the credible interval on each centroid measurement, we implemented the flux randomization/random subset selection (FR/RSS) method of \cite{Peterson+1998}, in which we performed $N=1,000$ iterations and measured the centroid each time. From this distribution of centroid measurements we then computed the $16^{\mathrm{th}}$ and $84^{\mathrm{th}}$ percentiles to estimate the $68~\mathrm{per~cent}$ credible interval for the observed centroid measurements.

\begin{table}
	\caption{Interpolated cross--correlation function centroids ($\tau_{\mathrm{cent}}$) relative to \textit{W2}. Negative values indicate a given band \textit{leads} the \textit{W2} data; positive values indicate a \textit{lag}. Parameter errors represent $68~\mathrm{per~cent}$ credible intervals and are shown to the precision of the \textsc{javelin} results for more direct comparison. A $\left(*\right)$ indicates that there was no reliable centroid measurement.}
	\centering
	\begin{tabular}{cccc}
		\hline
		Waveband & \multicolumn{3}{c}{$\tau_{\mathrm{cent}}$ [days]} \\
        & Unaltered & De--trended & \textsc{javelin} \\
		\hline
		\textit{X}  & $-0.39_{-0.08}^{+0.13}$  & $-0.38_{-0.07}^{+0.09}$  & $\left(*\right)$ \\
		\textit{W2} & $0.03_{-0.10}^{+0.07}$   & $-0.01_{-0.06}^{+0.05}$  & $0.00\pm0.01$ \\
		\textit{M2} & $0.14_{-0.11}^{+0.08}$   & $0.06_{-0.06}^{+0.08}$   & $0.03\pm0.02$ \\
		\textit{W1} & $0.21_{-0.08}^{+0.09}$   & $0.15\pm0.07$            & $0.13\pm0.03$ \\
		\textit{U}  & $0.32_{-0.08}^{+0.10}$   & $0.32_{-0.07}^{+0.08}$   & $0.29\pm0.04$ \\
		\textit{B}  & $0.29_{-0.15}^{+0.09}$   & $0.25_{-0.12}^{+0.13}$   & $0.24\pm0.06$ \\
		\textit{V}  & $0.22_{-0.25}^{+0.21}$   & $0.23_{-0.26}^{+0.24}$   & $0.20_{-0.11}^{+0.12}$ \\
		\hline
		\label{tab:centroids}
	\end{tabular}
\end{table}

\begin{figure}
	\begin{center}
		\scalebox{1.0}{\includegraphics[width=\linewidth]{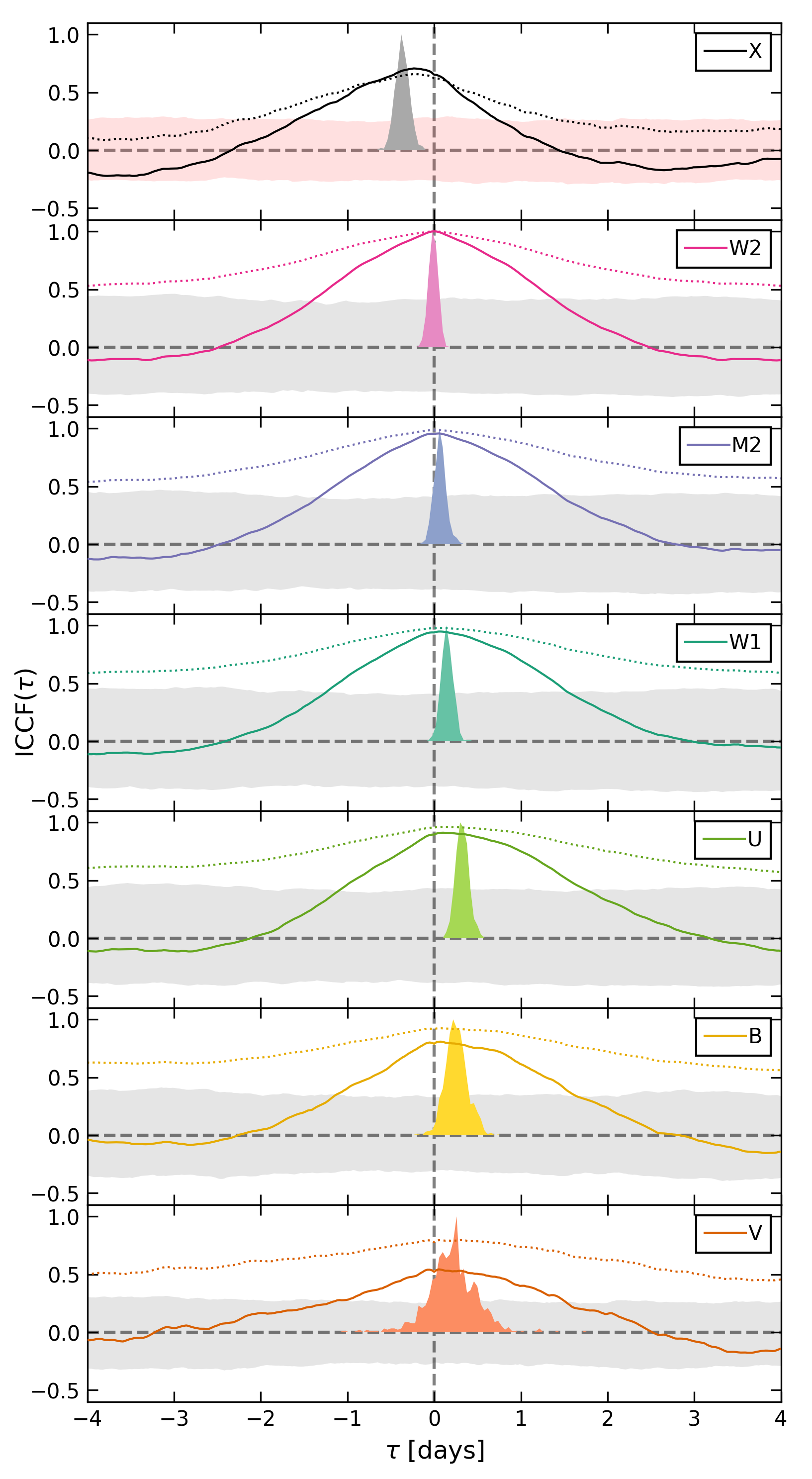}}
		\caption{Interpolated cross--correlation functions for the unaltered (dotted) and de--trended (solid) light curves in each filter relative to the \textit{W2} light curve. Negative $\tau$ values indicate a given band \textit{leads} the \textit{W2} data; positive values indicate a \textit{lag}. Colour--coded histograms represent the centroid distributions obtained from the $N=1,000$ FR/RSS evaluations of the de--trended ICCFs. Shaded horizontal regions are bound by the $1^{\mathrm{st}}$ and $99^{\mathrm{th}}$ percentiles of the $N=1,000$ light curve simulations (see text for details). The dashed lines are the zero--point for each axis.}
		\label{fig:ccfs}
	\end{center}
\end{figure}

The ICCFs and centroid distributions are shown in Figure \ref{fig:ccfs}, with centroid measurements and credible intervals given in Table \ref{tab:centroids}. We note that the use of unaltered versus de--trended light curves does not significantly change our results; de--trended centroid estimates produce credible intervals $\sim10~\mathrm{per~cent}$ smaller than the unaltered ones. As is visually evident in Figure \ref{fig:lightcurves}, all of the de--trended UV/optical light curves are found to be strongly correlated with the \textit{W2} data (i.e. $\max\left[\mathrm{ICCF}\left(\tau\right)\right]\gtrsim0.8$), with only the \textit{V} band data exhibiting a significantly weaker correlation strength ($\max\left[\mathrm{ICCF}\left(\tau\right)\right]\approx0.5$) and broader centroid distribution, likely due to its lower observed rapid variability. The X--rays exhibit a moderately strong correlation with the \textit{W2} data ($\max\left[\mathrm{ICCF}\left(\tau\right)\right]\approx0.7$), and while the peak of the X--ray ICCF indicates a modest \textit{lead} with respect to the \textit{W2} data of $\tau_{\mathrm{peak}}\approx0.2~\mathrm{d}$, due to the asymmetry of the peak its centroid is measured as a \textit{lead} of $\tau_{\mathrm{cent}}\approx0.4~\mathrm{d}$.

An alternative approach to estimating inter--band continuum lags is to model some driving light curve as a damped random walk which then becomes lagged, smoothed, and scaled in the various response light curves. This approach is implemented in the \textsc{javelin} code \citep{Zu+2013}, which we use here to compare with our ICCF centroid measurements by using our unaltered light curves\footnote{\textsc{javelin} performs its own de--trending procedure as part of its methodology.}. Our inter--band lag measurements with \textsc{javelin} were computed using its built--in MCMC sampler with $100$ walkers over $100,000$ iterations after burning the first $100,000$ iterations. Parameter values and $68~\mathrm{per~cent}$ credible intervals were estimated by taking the median as well as $16^{\mathrm{th}}$ and $84^{\mathrm{th}}$ percentiles of the corresponding posterior distribution. Our results are given in Table \ref{tab:centroids} where we find that, apart from producing credible intervals $\sim55~\mathrm{per~cent}$ smaller than the de--trended ICCF ones, the \textsc{javelin} results are completely consistent with our ICCF results. We note, however, that we are unable to place any constraint on the X--ray lag with \textsc{javelin} as it is extremely sensitive to the top--hat width of the response function implemented in the code, which we do not know \textit{a priori}.

For a standard accretion disc where $T \propto r^{-3/4}$ the lag--wavelength spectrum is expected to follow $\tau \propto \lambda^{4/3}$ as $\lambda \propto T^{-1}$ and $\tau = r/c$. Our results, however, exhibit a completely different behaviour as we find that the optical bands all exhibit a nearly constant \textit{lag} of $\sim0.3~\mathrm{days}$ relative to \textit{W2}. Furthermore, while the \textit{U} band lag may appear to follow the standard prediction, it is likely that the measured lag relative to \textit{W2} is enhanced due to significant contamination from the H$\beta$ emission line originating in the BLR, and is thus unlikely to be entirely due to disc reprocessing. We note that the optical flattening of the lag--wavelength spectrum was also observed by \cite{Troyer+2016} using a 2012 \textit{Swift} campaign. However, the low--cadence of that campaign ($75$ observations over $3$ months) resulted in much larger uncertainties on the measured inter--band lags (a factor of $\sim10$ larger than those presented here) and thus no conclusive evidence of a flattened lag--wavelength spectrum.

\subsection{Extracting the variable spectral component} \label{sec:fluxflux}
The fractional variability results in Table \ref{tab:obslog} decrease rapidly at longer wavelengths, likely due to the presence of a constant component that dominates the redder wavebands, such as emission from the host galaxy. To disentangle the variable AGN component from the invariable host galaxy emission, we implement a flux--flux analysis \citep[e.g.][]{McHardy+2018}.

We first extract the flux density light curve ($F_{\nu}\left(\lambda,t\right)$) in each UVOT filter and de--redden them according to a Galactic extinction of $E\left(B-V\right)=0.207$ \citep{Willingale+2013} using the \cite{Cardelli+1989} extinction law with $R_V=3.1$. We filtered the light curves for observations that used every UVOT filter (total of 217 points) and fit them using $\chi^{2}$ minimization with a linear trend line as:
\begin{equation} \label{eqn:fluxflux}
    F_{\nu}\left(\lambda,t\right) = A\left(\lambda\right)+R\left(\lambda\right)X\left(t\right),
\end{equation}
where $X\left(t\right)$ is a dimensionless driving light curve with mean of 0 and standard deviation of 1, $A\left(\lambda\right)$ is a wavelength--dependent shift factor, and $R\left(\lambda\right)$ is a wavelength--dependent scale factor representing the RMS spectrum. Here, $X\left(t\right)$ is simply derived by computing the mean of the normalized (subtracting the mean, dividing by the standard deviation) UVOT light curves\footnote{We tested allowing each point in $X\left(t\right)$ to be free--to--vary, finding no significant differences in interpretation of the final results.}. To isolate the intrinsic AGN variable component, we evaluate the best--fit line in each waveband at $\mathrm{max}\left[X\left(t\right)\right]$ and $\mathrm{min}\left[X\left(t\right)\right]$, and take the difference, $D\left(\lambda\right)$. To isolate the host galaxy component, we solve for $X\left(t\right)$ such that $F_{\nu}\left(\textit{W2},t\right)=0$, then extrapolate the best--fit line of each other waveband to this value to estimate the minimum host galaxy contribution in each, $G\left(\lambda\right)$.

\begin{table}
	\caption{Flux--flux fit parameters according to Equation \ref{eqn:fluxflux} and computed values to isolate the AGN ($D$) and host galaxy ($G$) components (see text for details), in units of mJy. The fit to all UVOT light curves yields $\chi^{2}_{\nu}=1.26$.}
	\centering
	\begin{tabular}{ccccc}
		\hline
		Waveband & \multicolumn{2}{c}{Fit Parameters} & \multicolumn{2}{c}{Computed Values} \\
        & $A\left(\lambda\right)$ & $R\left(\lambda\right)$ & $D\left(\lambda\right)$ & $G\left(\lambda\right)$\\
		\hline
  		\textit{W2} & $6.69\pm0.01$  & $1.71\pm0.02$ & $6.89\pm0.05$ & $-$ \\
		\textit{M2} & $6.46\pm0.02$  & $1.46\pm0.02$ & $5.89\pm0.06$ & $0.74\pm0.09$ \\
		\textit{W1} & $7.33\pm0.02$  & $1.35\pm0.02$ & $5.44\pm0.06$ & $2.05\pm0.09$ \\
		\textit{U}  & $6.37\pm0.01$  & $1.03\pm0.02$ & $4.17\pm0.05$ & $2.32\pm0.07$ \\
		\textit{B}  & $9.67\pm0.02$  & $0.82\pm0.02$ & $3.31\pm0.06$ & $6.46\pm0.09$ \\
		\textit{V}  & $15.49\pm0.04$ & $0.72\pm0.04$ & $2.91\pm0.12$ & $12.66\pm0.15$ \\
		\hline
		\label{tab:fluxfluxpars}
	\end{tabular}
\end{table}

\begin{figure}
	\begin{center}
		\scalebox{1.0}{\includegraphics[width=\linewidth]{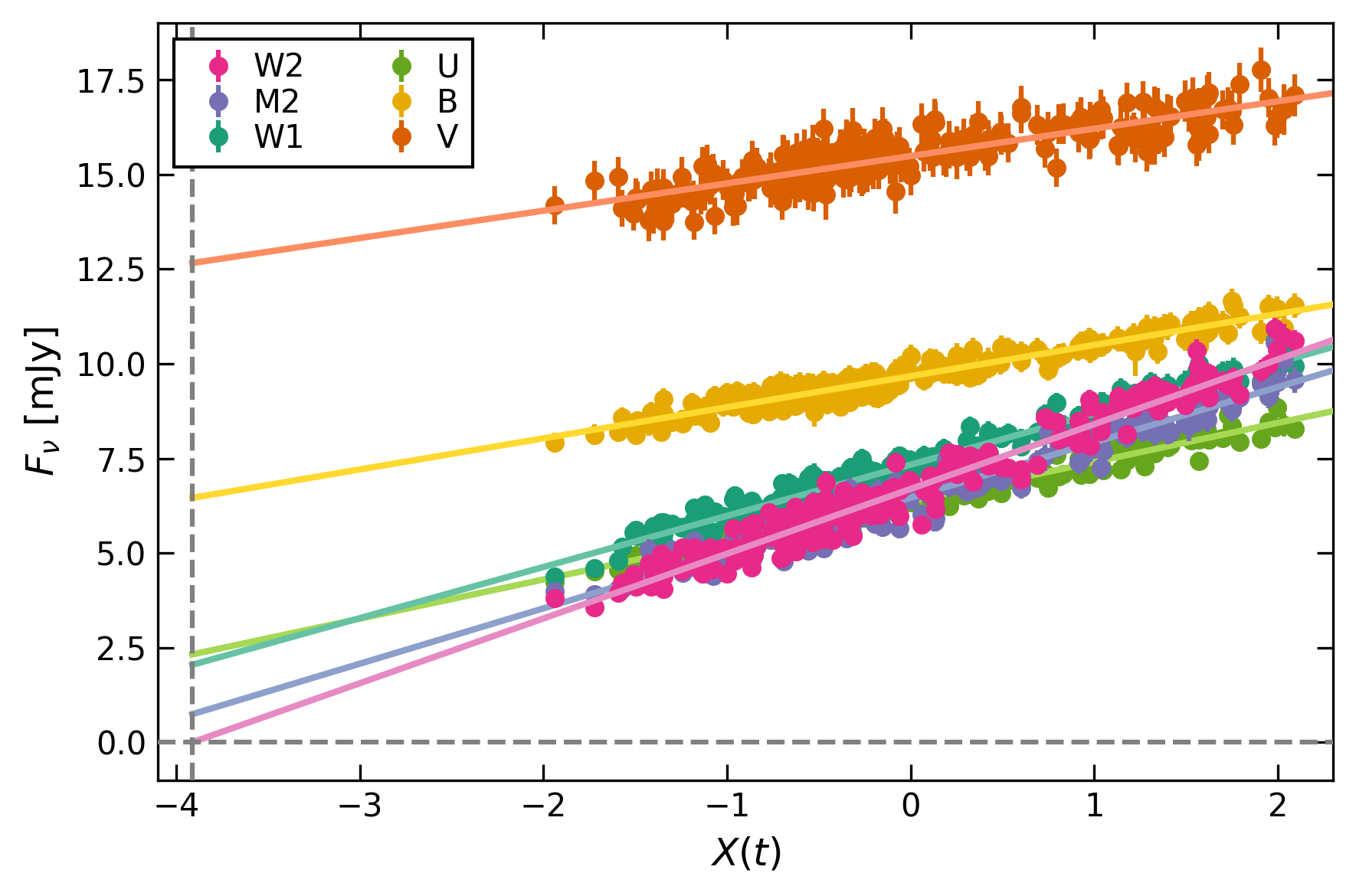}}
        \\
        \scalebox{1.0}{\includegraphics[width=\linewidth]{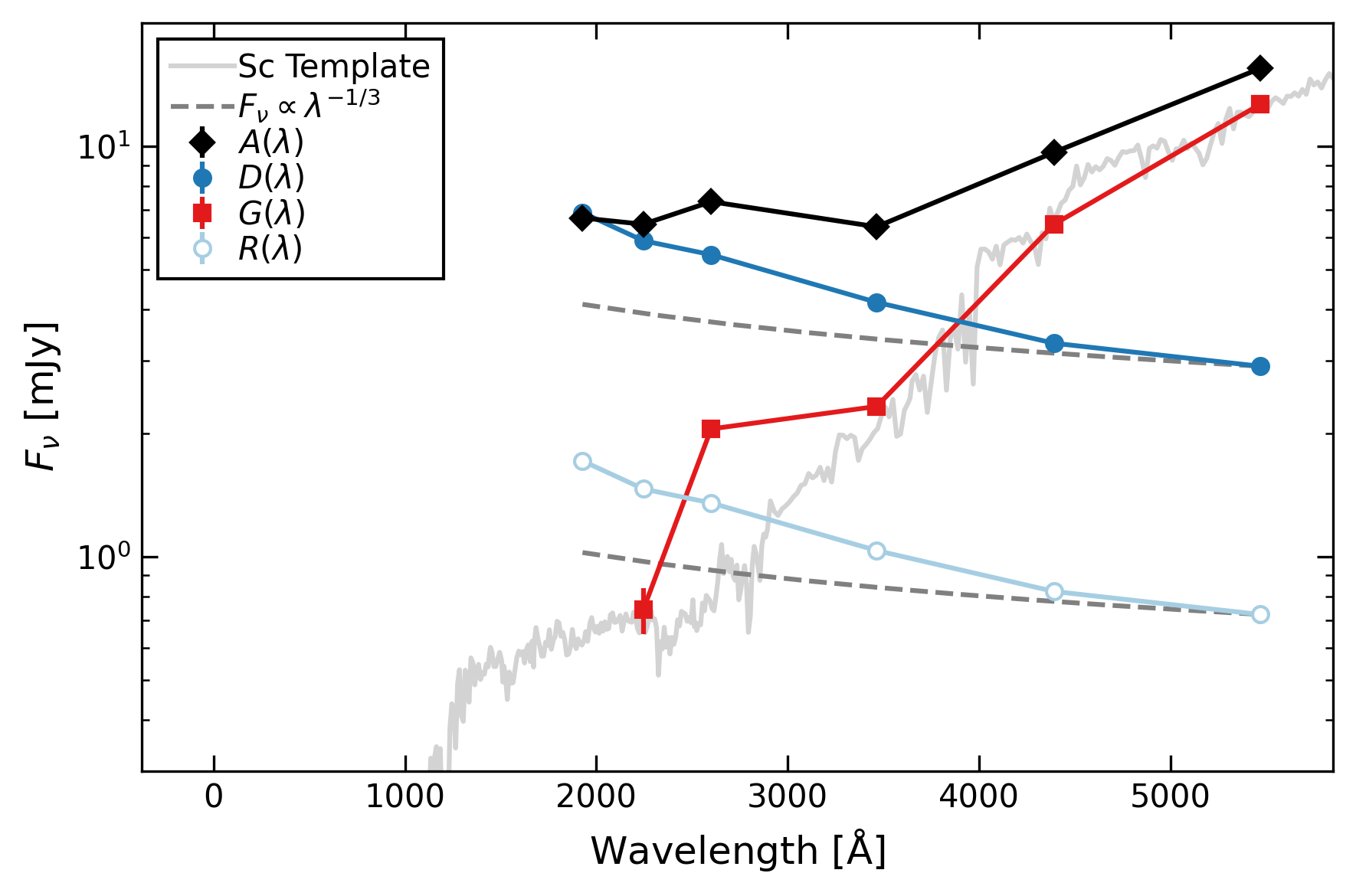}}
		\caption{\textit{Top}: Flux--flux plots for all de--reddened and filtered UVOT flux density light curves. Colour--coded solid lines represent the best--fit lines $F_{\nu}\left(\lambda,t\right)=A\left(\lambda\right)+R\left(\lambda\right)X\left(t\right)$, with fit parameters given in Table \ref{tab:fluxfluxpars}. The vertical grey dashed line represents the \textit{W2} zero crossing point, which is used to estimate the minimum host galaxy contribution in each other band. \textit{Bottom}: Measured total, $A\left(\lambda\right)$, and RMS, $R\left(\lambda\right)$, spectra alongside the computed AGN, $D\left(\lambda\right)$, and host galaxy, $G\left(\lambda\right)$, spectra based on the best--fit lines to the flux--flux plots in the top panel (all values given in Table \ref{tab:fluxfluxpars}). The expected accretion disc spectrum of $F_{\nu} \propto \lambda^{-1/3}$ is plotted alongside the RMS and AGN spectra, scaled by the \textit{V} band flux of each, as the dashed dark grey curves. The Sc spiral template is shown as the solid light grey curve.}
		\label{fig:fluxflux}
	\end{center}
\end{figure}

The flux--flux plots and best--fit lines are shown in Figure \ref{fig:fluxflux} (top panel), with parameters and corresponding computed values shown in Figure \ref{fig:fluxflux} (bottom panel) and given in Table \ref{tab:fluxfluxpars}. No obvious deviations from a linear trend are present in the data ($\chi^2_{\nu}=1.26$), thus suggesting no significant changes in the spectral shape during the campaign. The spectrum is dominated by host galaxy emission in the \textit{B} and \textit{V} bands, thus resulting in the rapid decrease in variability in those bands relative to the others. We fit the host galaxy emission using the spiral galaxy templates available in the SWIRE template library\footnote{\href{http://www.iasf-milano.inaf.it/~polletta/templates/swire\_templates.html}{http://www.iasf-milano.inaf.it/~polletta/templates/swire\_templates.html}} \citep{Polletta+2007}, finding the Sc spiral template to provide the best description of the data, consistent with the host galaxy morphological classification of NGC\,6814 as an SAB(rs)bc galaxy by \cite{deVaucouleurs+1991}. We note that the excess emission measured in the \textit{W1} band relative to the Sc spiral template may be due to the `small blue bump' feature consisting of blended Fe \textsc{II} and Balmer continuum emission \citep[see, e.g., Figure 5 of][]{Mehdipour+2015}. 

The intrinsic AGN variability spectrum does not follow a typical accretion disc spectrum, for which $F_{\nu} \propto \lambda^{-1/3}$, but instead appears extremely blue such that $F_{\nu} \propto \lambda^{-0.86}$. While we do not consider the effects of extinction in the source frame here, we note that its effect would be to make the intrinsic AGN variability spectrum even bluer and thus more extreme.

\section{Discussion} \label{sec:discussion}
In the previous sections, we have presented an analysis of the temporal properties of NGC\,6814 in the X--ray, UV, and optical wavebands from our 2022 \textit{Swift} campaign consisting of $\sim250$ observations over $75$ days. We have found that the long--term X--ray variability is distinctly different than the long--term UV/optical variability, as evidenced by its unique long--term trend (Figure \ref{fig:lightcurves}) and comparatively flat SF slope (Figure \ref{fig:structurefunctions} and Table \ref{tab:sfparams}). Short--term variations in the X--ray band were found to \textit{lead} the UV (\textit{W2}) ones by $\sim0.4~\mathrm{d}$ (Figure \ref{fig:ccfs} and Table \ref{tab:centroids}). The UV and optical wavebands all exhibit similar long--term trends (Figure \ref{fig:lightcurves}) and SF properties (Figure \ref{fig:structurefunctions} and Table \ref{tab:sfparams}), suggesting a common origin. Interestingly, the inter--band lags do not follow the $\tau\propto\lambda^{4/3}$ predicted relation of a standard disc reprocessing scenario (Table \ref{tab:centroids}), nor does the intrinsic AGN variability spectrum follow the $F_{\nu}\propto\lambda^{-1/3}$ predicted relation of a standard accretion disc (Figure \ref{fig:fluxflux} and Table \ref{tab:fluxfluxpars}). 

To probe the origin of these peculiar results, we use the \textsc{kynxiltr} (Kammoun et al. 2023, submitted) and \textsc{kynsed}\footnote{\href{https://projects.asu.cas.cz/dovciak/kynsed}{https://projects.asu.cas.cz/dovciak/kynsed}} \citep{Dovciak+2022} codes to self--consistently compute the inter--band lags and AGN SED, respectively, for a selection of physical parameter combinations. Both codes self--consistently compute the X--ray power law, reflected X--ray, as well as thermal disc emission spectra and response functions based on the following set of physical parameter inputs: black hole mass $M_{\mathrm{BH}}$, black hole spin $a^{*}$, line--of--sight inclination $i$ where $i=0$ is face--on and $i=90$ is edge--on, mass accretion rate $\dot{m}_{\mathrm{Edd}}$, colour--temperature correction factor $f_{\mathrm{col}}$, fraction of total accretion luminosity powering the corona $L_{\mathrm{transf}}$, outer disc radius $R_{\mathrm{out}}$, Fe abundance $A_{\mathrm{Fe}}$, X--ray coronal height $h$, X--ray photon index $\Gamma$, source redshift $z$, and model normalization $1/D_{\mathrm{L}}^2$ where $D_{\mathrm{L}}$ is the luminosity distance to the source. We note that here we have assumed the following standard values in all forthcoming computations: an inner disc radius of $R_{\mathrm{in}}=R_{\mathrm{ISCO}}$ where $R_{\mathrm{ISCO}}$ is the radius of the innermost stable circular orbit, a constant density throughout the disc of $n_{\mathrm{e}}=10^{15}~\mathrm{cm}^{-3}$, and a high energy cut--off for the X--ray power law of $E_{\mathrm{cut}}=300~\mathrm{keV}$. Below we describe our choice of parameter values, which are fixed to the listed values unless otherwise stated.

Based on the X--ray spectral fits of \cite{GalloGonzalezMiller2021}, we assume $a^{*}=0.998$, $A_{\mathrm{Fe}}=4$, and $\Gamma=2$. While those authors found $i=65^{\circ}$ to best fit the 2016 \textit{XMM-Newton} spectrum of NGC\,6814, we find that the time--averaged XRT spectrum of our \textit{Swift} monitoring campaign is best fit with $i=45^{\circ}$, and therefore use the latter value here. \cite{GalloGonzalezMiller2021} also estimated the size of the extended X--ray corona to be $\sim25~r_g$ in diameter. Here, we approximate such an extended coronal geometry under the lamp--post coronal geometry used by the \textsc{kynxiltr} and \textsc{kynsed} codes by assuming $h=10~r_g$, which is comparable to a lamp--post placed along the spin axis of the black hole at nearly the maximum radial extent of the extended coronal geometry. The SMBH mass in NGC\,6814 has been measured as $M_{\mathrm{BH}}=\left(1.85\pm0.35\right)\times10^{7}~M_{\odot}$ based on its H$\beta$ emission line properties \citep{Bentz+2009}. While significantly lower mass estimates exist \citep[e.g.][]{Pancoast+2014}, we find that such low SMBH masses consistently produce disc spectra that significantly over--predict the observed soft X--ray flux which cannot be reconciled even with extreme X--ray absorption, and thus we do not consider them further here. We assume $f_{\mathrm{col}}=1.7$, consistent with the colour--temperature correction factor of black hole X--ray binary accretion discs \citep[e.g.][]{Shimura+1995}, and note that while the precise value used does affect the other estimated disc parameters, the interpretation of our results does not change (see Appendix \ref{app:degeneratekynpars} for details). Lastly, we use $z=0.00522$ \citep{Springob+2005} and $D_{\mathrm{L}}=21.65~\mathrm{Mpc}$ \citep{Bentz+2019}.

Therefore, we have left $R_{\mathrm{out}}$ and $\dot{m}_{\mathrm{Edd}}$ as variable parameters to explore. Of all the parameters explored by \cite{Kammoun+2021_kynxilrev}, only the reduction of outer disc radius was found to produce significant flattening of the lag--wavelength spectrum in a way comparable to our lag measurements in the \textit{B} and \textit{V} bands. We take $R_{\mathrm{out}}=10,000~r_g$ as a fiducial value for this parameter to compare with other values sampled in our fits. The \textit{observed} accretion rate ($\lambda_{\mathrm{Edd}}$) may be estimated by estimating the bolometric luminosity ($L_{\mathrm{bol}}$) as $L_{\mathrm{bol}} \approx K_{\mathrm{X}} L_{2-10~\mathrm{keV}} = 3.1\times10^{43}~\mathrm{ergs~s^{-1}}$, where $K_{\mathrm{X}}=15.5$ is the bolometric correction factor estimated based on Equation 3 of \cite{Duras+2020} and $L_{2-10~\mathrm{keV}}\approx2\times10^{42}~\mathrm{ergs}~\mathrm{s}^{-1}$ is the unabsorbed X--ray luminosity of the time--averaged XRT spectrum, yielding a sub--Eddington \textit{observed} accretion rate of $\lambda_{\mathrm{Edd}}=L_{\mathrm{bol}}/L_{\mathrm{Edd}}\approx0.0133$. We use this \textit{observed} accretion rate as a reasonable approximation of the \textit{intrinsic} accretion rate (i.e. within a factor of a few), and therefore take $\dot{m}_{\mathrm{Edd}}=0.0133$ as a fiducial value for this parameter to compare with other values sampled in our fits.

\begin{figure*}
	\begin{center}
        \scalebox{1.0}{\includegraphics[width=\linewidth]{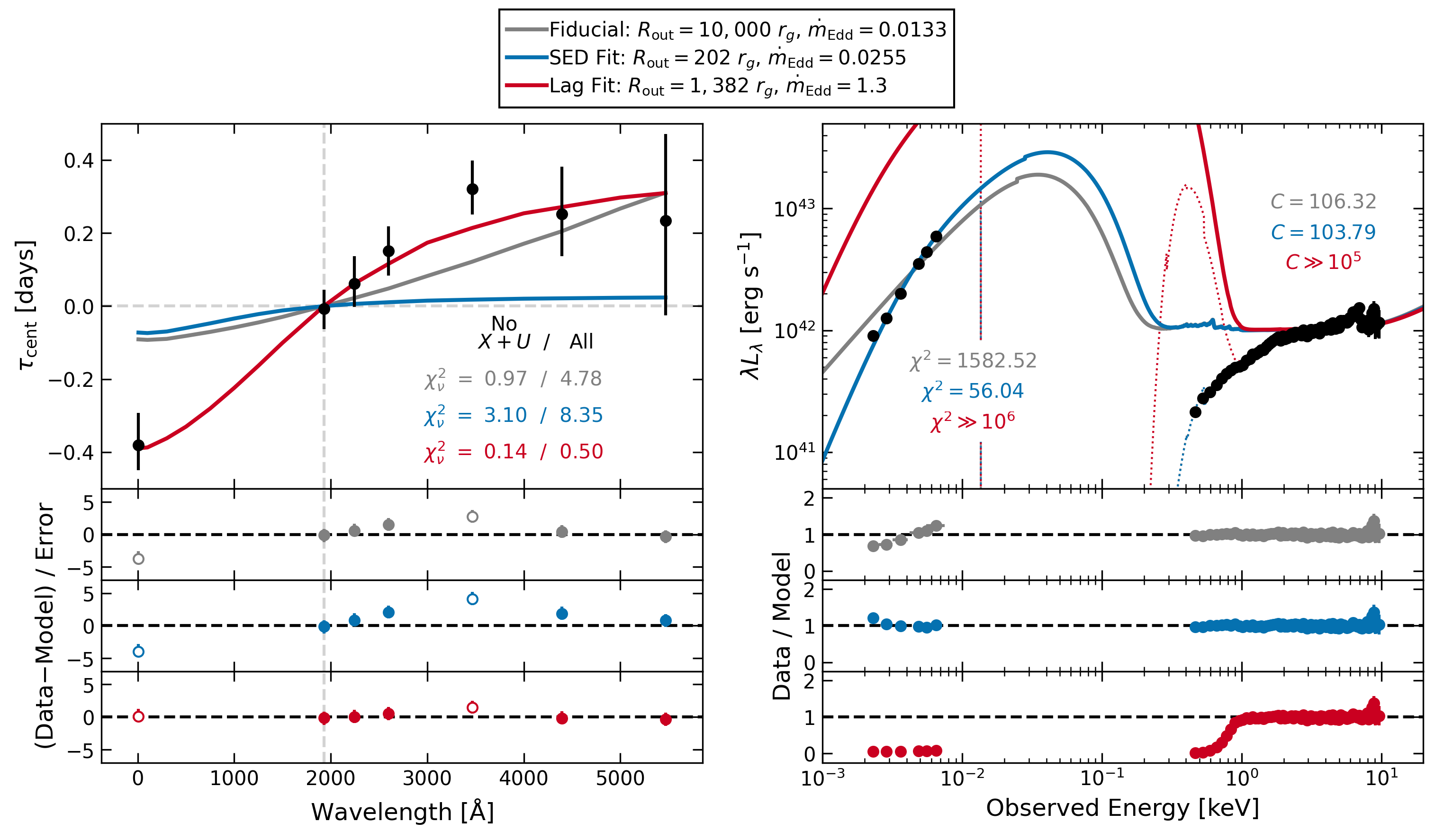}}
		\caption{The lag--wavelength spectrum (top left) and AGN SED (top right; observed as dotted curves, intrinsic as solid curves) are shown as the black data points. Each panel in the top row displays the results from individually fitting the lag--wavelength spectrum (red) and SED (blue) with \textsc{kynxiltr} and \textsc{kynsed}, respectively, as well as a fiducial (grey) curve based on appropriate parameter values for comparison. We include the relevant colour--coded fit statistics in each panel of the top row. For the lag fits, we have computed fit statistics by excluding the X--ray and \textit{U} band results as well as by including all of the data. For the SED fits, the fit statistics for the UV/optical ($\chi^2$) and X--ray ($C$) data are shown separately, near the corresponding data. The bottom three rows display the colour--coded best--fit model residuals corresponding to the fiducial, SED, and lag fits from top to bottom.}
		\label{fig:lagwavelength}
	\end{center}
\end{figure*}

We computed the inter--band lags and AGN SED with \textsc{kynxiltr} and \textsc{kynsed}, respectively, by fixing all aforementioned physical parameters to the given values while testing various combinations of $R_{\mathrm{out}}$ and $\dot{m}_{\mathrm{Edd}}$. In practice, we fit the observed lag--wavelength spectrum by computing the predicted inter--band lags in the XRT ($0.3-10~\mathrm{keV}$) and UVOT \citep[as listed in Table 1 of ][]{Edelson+2015} band--passes by stepping through $R_{\mathrm{out}}/r_g \in \left[55,~5,000\right]$ by $\Delta\log\left(R_{\mathrm{out}}/r_g\right)=0.01$ and through $\dot{m}_{\mathrm{Edd}} \in \left[0.003,~30\right]$ by $\Delta\log\left(\dot{m}_{\mathrm{Edd}}\right)=0.01$. We note that $L_{\mathrm{transf}}$ was fixed for each $\dot{m}_{\mathrm{Edd}}$ to ensure that the model predicted X--ray flux matched the observed one. For each $\left(R_{\mathrm{out}},\dot{m}_{\mathrm{Edd}}\right)$ pair, we evaluated the goodness of fit of the predicted lag values compared to the measured ones using the $\chi^2$ statistic. The \textit{U} band was excluded in all model evaluations due to possible contamination from the H$\beta$ emission line and/or diffuse continuum emission \cite[DCE; e.g.][]{KoristaGoad2019} originating in the BLR. While some physical scenarios, such as reprocessing between the X--ray and UV/optical emitting regions \citep[e.g.][]{McHardy+2018}, may produce X--ray lags (relative to the UV/optical) significantly different than those predicted by a purely disc reprocessing scenario, we found that exclusion of the X--ray lag value during model evaluation resulted in an extremely poor constraint on the accretion rate; therefore we included the X--ray lag measurement in all of our model evaluations. The intrinsic AGN variability spectrum\footnote{The data were converted into PHA and RSP files readable by \textsc{xspec} using the \textsc{ftflx2xsp} \textsc{ftools} task.} and optimally binned\footnote{The data were binned using the \textsc{grouptype=opt} flag of the \textsc{ftgrouppha} \textsc{ftools} task.} \citep{KaastraBleeker2016} time--averaged XRT spectrum were jointly fit with \textsc{kynsed} using \textsc{xspec} version 12.12.1 \citep{Arnaud1996} by computing the $\chi^2$ statistic for the intrinsic AGN variability spectrum and the Cash statistic \citep[i.e. $C$;][]{Cash1979} for the XRT spectrum to evaluate the model likelihood. Here, $L_{\mathrm{transf}}$ was left free--to--vary in order to fit the observed X--ray flux. We note that the $0.3-2~\mathrm{keV}$ band exhibits evidence of significant neutral absorption in the source frame. This was fit using \textsc{partcov*zphabs} in \textsc{xspec}, which we found to have a column density of $N_{\mathrm{H}}\approx10^{22}~\mathrm{cm}^{-2}$ and covering fraction of $f_{\mathrm{cov}}\approx0.4$, which were fixed throughout all of our fits. Galactic absorption was fit using \textsc{phabs} with $N_{\mathrm{H}}=1.53\times10^{21}~\mathrm{cm}^{-2}$ \citep{Willingale+2013}. Since the intrinsic AGN variability spectrum was already de--reddened for Galactic extinction (see Section \ref{sec:fluxflux}), we do not need to account for it in our \textsc{xspec} model. We recorded the parameters that best reproduced the inter--band lags and AGN SED individually, and computed the predicted corresponding curves, as well as computed a pair of fiducial curves based on $R_{\mathrm{out}}=10,000~r_g$ and $\dot{m}_{\mathrm{Edd}}=0.0133$. Error bars on forthcoming parameter values represent corresponding $68~\mathrm{per~cent}$ parameter credible intervals. 

Our results are shown in Figure \ref{fig:lagwavelength}. We find that the measured lag--wavelength spectrum is best fit by a moderately truncated outer disc radius of $R_{\mathrm{out}}=1,382^{+398}_{-404}~r_g$ at an accretion rate of $\dot{m}_{\mathrm{Edd}}=1.3^{+2.1}_{-0.9}$. This model self--consistently explains the X--ray lag, requiring no extended reprocessor scenario, and exhibits only a minor deviation in the \textit{U} band ($1.4\sigma$), suggesting only a marginal contribution from DCE. The corresponding predicted AGN SED, however, over--predicts the observed one by a factor of $\sim10$ across the UV/optical bands, and even extends well into the soft X--ray band where it over--predicts the observed $0.3-1~\mathrm{keV}$ flux by a factor of $\sim20$. In contrast, we find that the AGN SED is best fit by an extremely truncated outer disc radius of $R_{\mathrm{out}}=202\pm5~r_g$ at a modest accretion rate of $\dot{m}_{\mathrm{Edd}}=0.0255\pm0.0006$ that is comparable to our fiducial estimate. The corresponding predicted lag--wavelength spectrum, however, significantly under--predicts the measured one in the \textit{W1}, \textit{U}, and \textit{B} bands, suggesting a significant contribution from DCE, and cannot self--consistently explain the X--ray lag, suggesting the presence of an extended X--ray reprocessor, providing a statistically unacceptable fit even when ignoring the X--ray and \textit{U} band lags. Our fiducial model produces a statistically acceptable fit to the lag--wavelength spectrum if we consider that the X--ray and \textit{U} band lags have origins other than due to disc reprocessing, however, the corresponding predicted disc spectrum provides a statistically unacceptable fit to the intrinsic AGN variability spectrum, thereby strongly disfavouring a standard disc scenario. It is therefore plausible that a standard disc illuminated by a lamp--post X--ray corona cannot self--consistently explain the observed properties of NGC\,6814 given the fact that none of the above results adequately fit both the inter--band lags and AGN SED simultaneously.

One possible mechanism through which the outer disc radius may become truncated is via dust formation in the disc once the temperature falls below the dust sublimation temperature. \cite{BaskinLaor2018} showed that the minimum radius for which this occurs is:
\begin{equation} \label{eqn:dust}
R_{\mathrm{dust}}=0.018 L_{\mathrm{opt,45}}^{1/2}~\mathrm{pc},
\end{equation}
where $L_{\mathrm{opt,45}} = \lambda L_{\lambda}$ at $4861~\text{\AA}$ in units of $10^{45}~\mathrm{ergs}~\mathrm{s}^{-1}$. For $R_{\mathrm{out}}=202~r_g$ and $\dot{m}_{\mathrm{Edd}}=0.0255$ we find that $L_{\mathrm{opt}} \approx 9.6 \times 10^{41}~\mathrm{ergs}~\mathrm{s}^{-1}$ yielding $R_{\mathrm{dust}} \approx 628~r_g$, while for $R_{\mathrm{out}}=1,382~r_g$ and $\dot{m}_{\mathrm{Edd}}=1.3$ we find that $L_{\mathrm{opt}} \approx 1.8 \times 10^{43}~\mathrm{ergs}~\mathrm{s}^{-1}$ yielding $R_{\mathrm{dust}}\approx2,700~r_g$. Both of these estimates place the minimum dust sublimation radius far outside the corresponding outer disc truncation radius, thus it seems unlikely that dust formation in the disc is responsible for truncation of the outer disc here.

Outer disc truncation may also be achieved through disc self--gravity. According to Equation 15 of \cite{LaorNetzer1989}, the self--gravity radius of a \cite{ShakuraSunyaev1973} accretion disc occurs at:
\begin{equation} \label{eqn:selfgravity}
    R_{\mathrm{sg}} = 2150~m_{9}^{-2/9} \dot{m}_{\mathrm{Edd}}^{4/9} \alpha_{\nu}^{2/9},
\end{equation}
where $m_{9}=M_{\mathrm{BH}}/10^{9}~M_{\odot}$ and $\alpha_{\nu}$ is the $\alpha$--disc viscosity parameter. Assuming $\alpha_{\nu}\approx0.02$ \citep[e.g.][]{Mishra+2016}, for $\dot{m}_{\mathrm{Edd}}=0.0255$ we find $R_{\mathrm{sg}}\approx428~r_g$ while instead for $\dot{m}_{\mathrm{Edd}}=1.3$ we find $R_{\mathrm{sg}}\approx2,458~r_g$. Both of these estimates are significantly larger than the corresponding outer disc truncation radius, thus it seems that self--gravity is also unlikely to be responsible for truncation of the outer disc here. 

The measured SF break--times of the de--trended UV/optical light curves were all found to be consistent with $\langle\tau_{\mathrm{break}}\rangle\approx2.30~\mathrm{d}$. We may determine the most plausible origin in the disc of this characteristic time--scale as due to light--crossing ($t_{\mathrm{lc}}=r/c$), dynamical ($t_{\mathrm{dyn}}=\sqrt{r^3/GM_{\mathrm{BH}}}$), thermal ($t_{\mathrm{th}} \approx t_{\mathrm{dyn}}/\alpha_{\nu}$), or viscous ($t_{\mathrm{vis}} \approx t_{\mathrm{th}}/\alpha_{\mathrm{H}}^{2}$, where $\alpha_{\mathrm{H}}=H/R\approx0.01$ is the disc aspect ratio) variability time--scales. Solving each relation for the radius that corresponds to $t=2.30~\mathrm{d}$ finds $R_{\mathrm{lc}}\approx2180~r_g$, $R_{\mathrm{dyn}}\approx168~r_g$, $R_{\mathrm{th}}\approx12.4~r_g$, and $R_{\mathrm{vis}}\ll1~r_g$. We may immediately rule out both a light--crossing origin, as it exceeds both of our estimated disc sizes, and a viscous origin, as it is unphysically small (i.e. within the event horizon). 

Regarding a dynamical origin, we note that the corresponding radius ($R_{\mathrm{dyn}}\approx168~r_g$) lies well within $R_{\mathrm{out}}=1,382~r_g$ for the $\dot{m}_{\mathrm{Edd}}=1.3$ disc scenario, and is nearly coincident with $R_{\mathrm{out}}=202~r_g$ for the $\dot{m}_{\mathrm{Edd}}=0.0255$ disc scenario. In the latter scenario, the long--term variability of the UV/optical emission is driven by dynamical variations originating very close to the outer edge of the accretion disc. This may be plausible in the context of a bowl--like disc geometry with a steep rim \citep[e.g.][]{Starkey+2023}, in which the outer regions of the disc are elevated and tilted toward our line--of--sight and therefore dominate the observed area of the disc. Moreover, these outer regions of the disc will then dominate the measured UV/optical inter--band lags such that they are significantly larger than predicted by a thin, flat disc geometry \citep[see Figure 5 of][]{Starkey+2023}, perhaps in a manner comparable to those measured here. For accretion discs that are misaligned with respect to the SMBH spin axis, torques due to Lense--Thirring precession can effectively warp (for low degrees of misalignment), or even tear apart (for high degrees of misalignment), the disc \citep[e.g.][]{Nixon+2012,DydaReynolds2020}, a phenomenon that may be even more prevalent in the case of binary SMBHs \citep{Nixon+2013}, thereby revealing more of the outer disc region than in a thin, flat disc geometry. Therefore, whether intrinsically bowl--like or warped / torn to appear as such, non--standard disc geometries may offer a self--consistent description of the inter--band lags, AGN SED, and characterisitc variability time--scales observed in NGC\,6814. This will be explored in more detail in a future work.

Regarding a thermal origin, we note that the corresponding radius ($R_{\mathrm{th}}=12.4~r_g$) is exactly coincident with the outer edge of the extended X--ray corona measured by \citep{GalloGonzalezMiller2021}, where a diameter of $\sim25~r_g$ was measured yielding $R_{\mathrm{cor}}\approx12.5~r_g$. If this extended corona exists between the SMBH event horizon and inner disc radius such that $R_{\mathrm{in}}=R_{\mathrm{cor}}$ \citep[similar to the geometry invoked by, e.g.,][]{Done+2012,KubotaDone2018}, it is possible that the UV/optical SFs are measuring thermal variations at the inner edge of the accretion disc. Furthermore, we note that the measured X--ray SF break--time ($\tau_{\mathrm{break}}=1.7^{+0.5}_{-0.4}~\mathrm{d}$) is broadly consistent with the mean UV/optical break--time ($\langle\tau_{\mathrm{break}}\rangle\approx2.30~\mathrm{d}$), possibly suggesting an emission region of similar size (i.e. at or within the inner disc radius). The difference between the measured X--ray and UV/optical SF slopes may then be the result of the different physical processes responsible for the production of the coronal and disc emission, respectively.

Our findings here make it abundantly clear then that NGC\,6814 exhibits complex multi--wavelength variability. The extreme discrepancy found when comparing model predictions by fitting the inter--band lag spectrum and AGN SED individually as well as the essentially wavelength--independent characteristic variability time--scales across all observed band--passes may, however, be indicative of a non--standard accretion disc geometry such as a bowl--like or warped/torn disc. NGC\,6814 is a prime target for multi--wavelength follow--up across a wider waveband to more fully investigate the peculiar and interesting properties presented here.

\section{Conclusions} \label{sec:conclusions}
We have presented a thorough timing analysis of the X--ray, UV, and optical emission of NGC\,6814 from our 2022 \textit{Swift} monitoring campaign comprised of $\sim250$ observations over $75$ days. 

The long--term X--ray variability displays significantly different properties than the long--term UV/optical variability as seen in the smoothed long--term light curve trends of each waveband. 

A SF analysis reveals that the UV/optical variability is likely driven by a common physical mechanism, while the X--ray variability likely driven by a separate mechanism. We find that the similar SF break--times observed across all wavebands are consistent with either dynamical variations originating near the outer edge of a highly truncated outer disc or thermal variations originating at the inner edge of a truncated inner disc. 

A cross--correlation analysis evaluating inter--band lags finds that short--term variations in the X--ray band \textit{lead} the UV ones by $\sim0.4~\mathrm{d}$ while the optical wavebands exhibit a constant \textit{lag} of $\sim0.3~\mathrm{d}$ relative the UV. These measurements do not follow the standard disc X--ray reprocessing prediction (i.e. $\tau \propto \lambda^{4/3}$), unless an extreme accretion rate ($\dot{m}_{\mathrm{Edd}}=1.3^{+2.1}_{-0.9}$) and moderately truncated outer disc radius ($R_{\mathrm{out}}=1,382^{+398}_{-404}~r_g$) are invoked. 

A flux--flux analysis reveals a constant, red component (i.e. the invariable host galaxy emission) as well as a variable, extremely blue ($F_{\nu} \propto \lambda^{-0.9}$) component (i.e. the intrinsic AGN variable emission). The variable component does not follow the standard disc spectrum (i.e. $F_{\nu} \propto \lambda^{-0.9}$), unless an extremely truncated out disc radius ($R_{\mathrm{out}}=202\pm5~r_g$) and modest accretion rate ($\dot{m}_{\mathrm{Edd}}=0.0255\pm0.0006$) are invoked.

Our results suggest the possibility of a non--standard disc geometry in NGC\,6814 that will be explored in a future work. These results also make a strong case for future studies including multi--wavelength follow--up across a wider combined band--pass to more thoroughly investigate the origin of the various measured peculiarities.

\section*{Acknowledgements}
This work made use of data supplied by the UK Swift Science Data Centre at the University of Leicester. This research has made use of the NASA/IPAC Extragalactic Database (NED), which is funded by the National Aeronautics and Space Administration and operated by the California Institute of Technology. EK acknowledges financial support from the Centre National d’Etudes Spatiales (CNES).

\section*{Data Availability}
The data presented here is publicly available through the NASA HEASARC Archive (https://heasarc.gsfc.nasa.gov/docs/archive.html) and \textit{Swift} observatory (https://www.swift.ac.uk/index.php) websites. All data and analysis tools are available vai the corresponding author (AGG) upon reasonable request.



\bibliographystyle{mnras}
\bibliography{references} 




\appendix

\section{Determination of the optimal smoothing filter width} \label{app:detfiltwidth}

In order to isolate the short--term variations, which may be associated with the reprocessing of X--rays in the accretion disc, from long--term variations, which may be due to changes in the behaviour of the accretion flow itself, we de--trended the light curves in the following way. We first computed the auto--correlation function (ACF) of the \textit{W2} light curve (this is the reference band used in the cross--correlation analysis) using the interpolated cross--correlation function (ICCF; \citealt{GaskellSparke1986,GaskellPeterson1987,WhitePeterson1994}; the details of our ICCF methodology are presented in Section \ref{sec:iccf}). We then smoothed the data using a Savitzky--Golay filter for a range of filter widths $w/\mathrm{d} \in \left[3, 65\right]$ in steps of $\Delta \left(w/\mathrm{d}\right)=0.5$, approximately. We note, however, that since the Savitzky--Golay filter actually smooths over a specific number of data points, and given the gappy, uneven sampling of the data which results in irregularly changing time widths for the moving filter with a set number of data points, we give the nearest approximate value for all related filter widths, which are within $\pm0.25~\mathrm{d}$ of their true values. For each filter width, we computed the \textit{W2} ACF and evaluated its standard deviation in the regions outside of the central peak at $\tau=0~\mathrm{d}$ (i.e. for $\left|\tau\right|$ greater than the first zero crossing points of the ACF peak). Minimizing this quantity minimizes the long--term variations present in the light curve not due to the thermal reprocessing of X--rays in the accretion disc that act to broaden the \textit{W2} ACF, and therefore all ICCFs that use it as a reference band, thereby enabling a more accurate and precise evaluation of the ICCF centroids. 

\begin{figure}
	\begin{center}
		\scalebox{1.0}{\includegraphics[width=\linewidth]{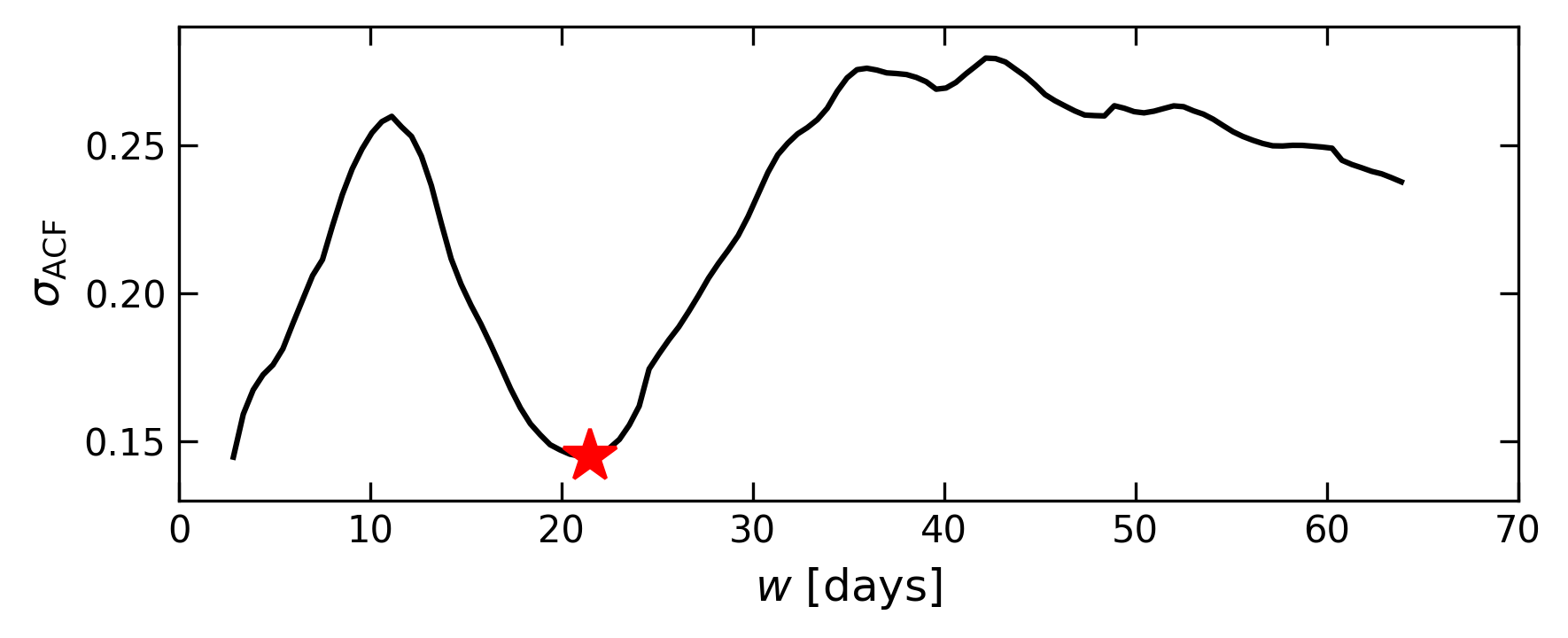}}
        \\
        \scalebox{1.0}{\includegraphics[width=\linewidth]{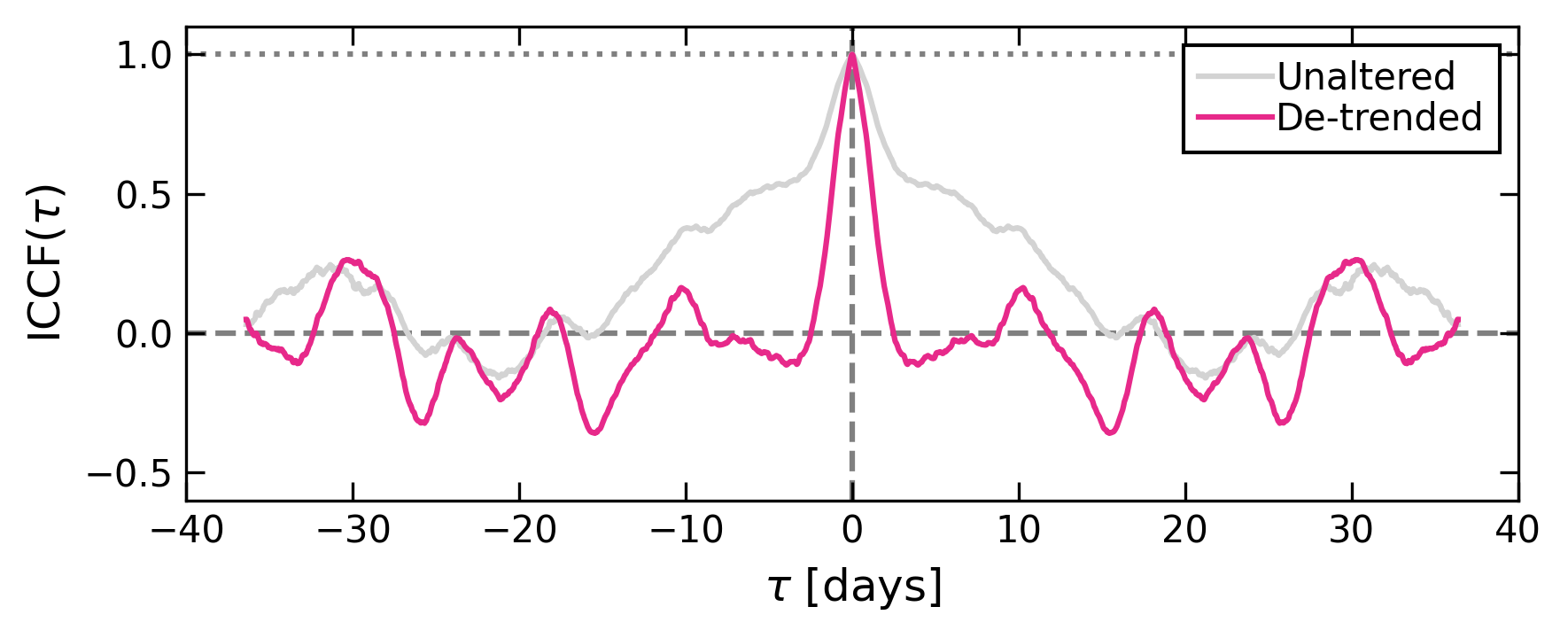}}
		\caption{\textit{Top}: The optimal filter width determination method (see text for details). The $21.5~\mathrm{d}$ filter width is shown as the red dot, which minimizes the standard deviation of the ACF outside the central peak at $\tau=0~\mathrm{d}$. \textit{Bottom}: The unaltered (grey) and de--trended (coloured) \textit{W2} ACFs, using a Savitzky--Golay filter width of $21.5~\mathrm{d}$ for the latter. The dashed lines are the zero--point for each axis, and the horizontal dotted line is $\mathrm{ACF}=1$.}
		\label{fig:w1acf}
	\end{center}
\end{figure}

The results of this procedure are shown in Figure \ref{fig:w1acf}, where we find an optimal Savitzky--Golay filter width of $21.5~\mathrm{d}$. We note that while the non--peak variance can be minimized further by using the narrowest filter widths tested, such smoothing also removes short--term variations that we are interested in evaluating as the smoothed trend essentially replicates the data.

\section{Colour--temperature correction factor and outer disc radius degeneracy} \label{app:degeneratekynpars}

The colour--temperature correction factor ($f_{\mathrm{col}}$) for AGN accretion discs is not well known and is likely to be different from that of black hole X--ray binary accretion discs, where $f_{\mathrm{col}}\sim1.7$ \citep{Shimura+1995}, due to the significantly lower peak temperatures (i.e. $T_{\mathrm{max}}\sim10^5~\mathrm{K}$ in AGN discs versus $T_{\mathrm{max}}\sim10^7~\mathrm{K}$ in black hole X--ray binary discs). Importantly, increasing values of colour--temperature correction factor shift the peak of the disc black body spectrum toward shorter wavelengths, resulting in reduced emission at longer wavelengths. Since the outer regions of the accretion disc are responsible for producing the emission at longer wavelengths, the reduction of this emission can be accommodated by increasing the size of the disc. Clearly, then, a degeneracy between color--temperature correction factor and outer disc radius is apparent. 

We therefore performed a number of fits to the AGN SED with \textsc{kynsed} \citep{Dovciak+2022} by fixing the colour--temperature correction factor at values of $f_{\mathrm{col}} \in \left[1,2.5\right]$ spaced by $\Delta f_{\mathrm{col}} = 0.1$, leaving $R_{\mathrm{out}}$, $\dot{m}_{\mathrm{Edd}}$, and $L_{\mathrm{transf}}$ as free--to--vary parameters and all other parameters (and model components) fixed to their listed values in Section \ref{sec:discussion}. For each fit, we computed the corrected Akaike Information Criterion \citep[AICc;][]{Akaike1974,HurvichTsai1989}, and determined the range of statistically equivalent (at the $95~\mathrm{per~cent}$ confidence level) values of colour--temperature correction factor to be those with $\exp\left[\left(\mathrm{AICc}_{\mathrm{min}}-\mathrm{AICc}_{i}\right)/2\right]\geq0.05$, where $\mathrm{AICc}_{\mathrm{min}}$ is the AICc value corresponding to the best fitting model and $\mathrm{AICc}_{i}$ is the AICc value corresponding to the $i^{\mathrm{th}}$ model realization.

\begin{figure}
	\begin{center}
		\scalebox{1.0}{\includegraphics[width=\linewidth]{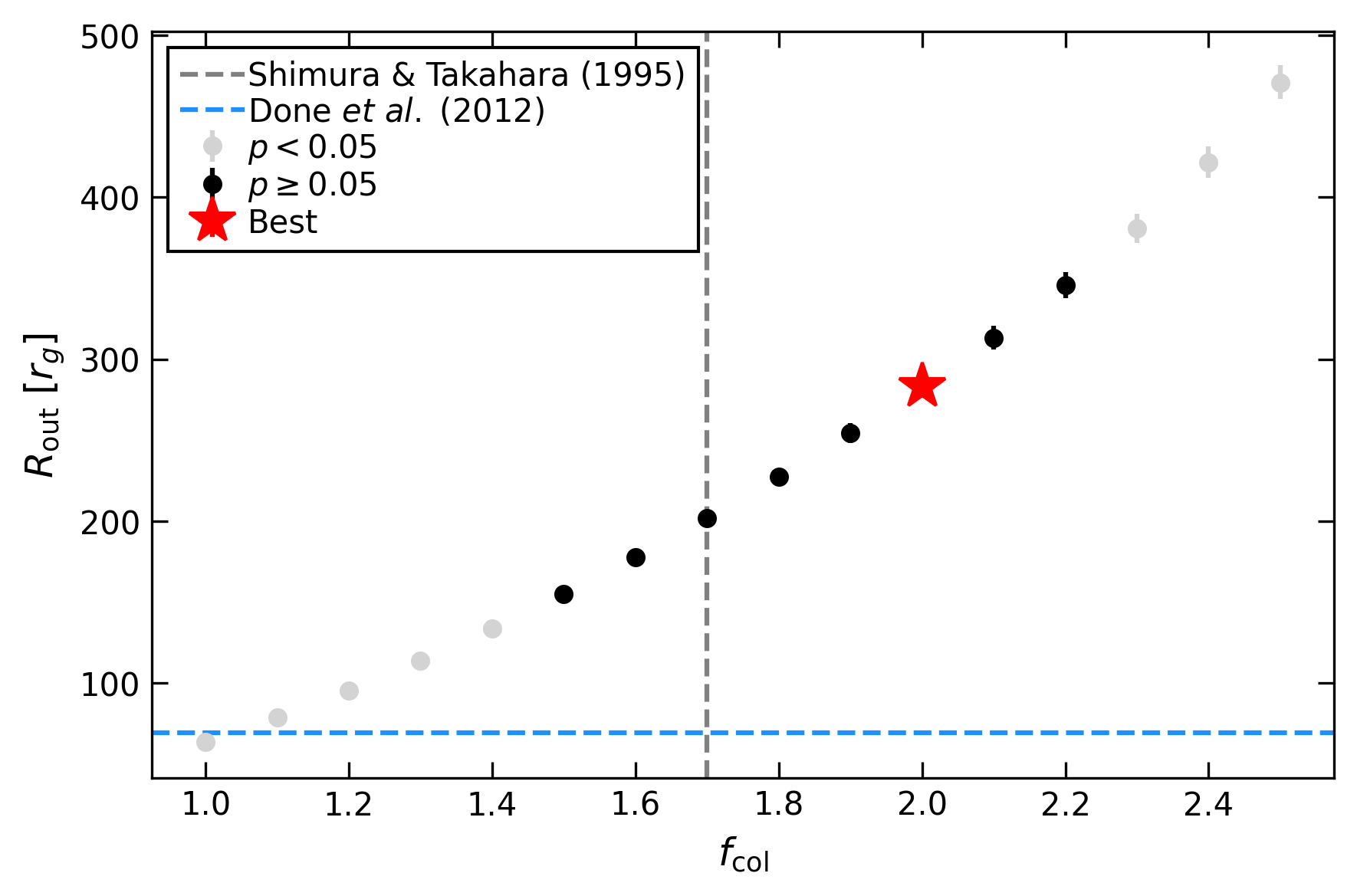}}
		\caption{Outer disc radius ($R_{\mathrm{out}}$) measurements for a sequence of fixed colour--temperature correction factors ($f_{\mathrm{col}}$). The red star represents the best overall fit to the data, while black (grey) points are statistically equivalent (ruled out) at the $95~\mathrm{per~cent}$ confidence level. The vertical black dashed line represents $f_{\mathrm{col}}=1.7$, which was measured by \protect\cite{Shimura+1995} in the accretion discs of black hole X--ray binaries. The horizontal blue dashed line represents the $R_{\mathrm{out}}$ measurement when computing $f_{\mathrm{col}}$ according to the \protect\cite{Done+2012} prescription.}
		\label{fig:fcolrout}
	\end{center}
\end{figure}

The results of this procedure are shown in Figure \ref{fig:fcolrout}. We find that the data are best fit with $f_{\mathrm{col}}=2$ resulting in $R_{\mathrm{out}}=283\pm7~r_g$, with an acceptable / statistically equivalent range of values being $f_{\mathrm{col}} \in \left[1.5,2.2\right]$ resulting in $R_{\mathrm{out}} \in \left[155,345\right]$. It is clear that despite the significant differences between AGN and black hole X--ray binary accretion discs, assuming $f_{\mathrm{col}}=1.7$ does not significantly affect our results as the data require significant truncation of the outer disc radius in all cases. We also attempted to fit the data using the temperature--dependent color--temperature correction factor for AGN discs developed by \cite{Done+2012}, which resulted in $R_{\mathrm{out}}=69\pm1~r_g$, thus requiring even more extreme outer disc truncation. Therefore, our fits presented in Section \ref{sec:discussion} had $f_{\mathrm{col}}=1.7$ fixed.


\bsp	
\label{lastpage}
\end{document}